\documentclass[aps,prl,showpacs,floatfix,twocolumn,superscriptaddress,amsmath,amssymb]{revtex4}

\usepackage{graphicx}
\usepackage{bm}
\usepackage{multirow}

\begin{document}

\title{Crowding induced entropy-enthalpy compensation in protein association equilibria}

\author{Young~C.~Kim}
\email{yckim@dave.nrl.navy.mil}
\affiliation{Center for Computational Materials Science, Naval
  Research Laboratory, Washington DC 20375, USA}
\author{Jeetain Mittal}
\email{jeetain@lehigh.edu}
\affiliation{Department of Chemical Engineering,
  Lehigh University, Bethlehem, PA 18015}

\begin{abstract}
A statistical mechanical theory is presented to predict the effects of
macromolecular crowding on protein association equilibria, accounting
for both excluded volume and attractive 
interactions between proteins and crowding molecules. Predicted
binding free energies are in excellent agreement with simulation
data over a wide range of crowder sizes and packing fraction. It is
shown that
attractive interactions between proteins and crowding agents
counteract the stabilizing effects of excluded volume interactions. A
critical attraction strength, for
which there is no net effect of crowding, is approximately independent of the crowder packing
fraction.

\end{abstract}

\maketitle

Protein-protein interactions are important in many essential
biological functions, such as transcription, translation, and signal
transduction~\cite{Kleanthous-2000}. A lot of progress has been made
in understanding protein association in dilute solution 
via experiments and simulations~\cite{Tang-N-2006,boehr2008,elcock1999,gilson2007}. 
Cells, on the other hand, contain various macromolecules, e.g., DNA, RNA, proteins,
organelles, etc., which constitute up to 40\% of the cell
volume~\cite{Fulton:1982p8511}. It is thus crucial to relate \textit{in vitro} experimental or simulation
results to those in a crowded cellular environment~\cite{Zimmerman:1993p2187,Cheung-2005,Zhou-ARB-2008,Shen-2009,Kim-BPJ-2009,Elcock-COSB-2010,mittal-2010}. 

Several experimental studies have been performed to 
understand protein-protein interactions in a crowded environment~\cite{minton-1981a,Jarvis-1990,berg-1999,Wenner:2008p8635,Morar:2001p8850,patel2002,kozer-2004,Zorrilla:2004p8683,Phillip-BPJ-2009,Wang-Biochemistry-2011,fodeke2011}.
Most attention has been paid to the 
steric excluded volume effects of
inert crowding agents on the formation of protein complexes~\cite{Minton-1983,Zhou:2004p4459,Kim-JCP-2010}. Very recent studies
have also started to probe 
the effects of attractive interactions between proteins and
crowders on protein association~\cite{Douglas-PRL-2009,Jiao-BPJ-2010,Rosen-JPC-2011,Phillip-PNAS-2012}. 
These studies have highlighted the importance of accounting for enthalpic effects 
arising from attractive interactions in addition to commonly invoked excluded volume effects. 
It was found that the enthalpic effects can actually 
increase the binding free energy (thereby destabilizing the bound complex) in contrast to 
predictions based on available theoretical models that can only capture entropic effects. 

Most theoretical models of crowding are based on scaled particle theory (SPT) of hard-sphere 
fluids~\cite{Lebowitz-JCP-1964} or its modified
versions and have been applied to interpret experimental and computational
results with varying success. The failure of these models in several
situations highlights an important role played by attractive crowder-protein interactions. 
In our earlier work~\cite{Rosen-JPC-2011}, we had proposed an {\em ad hoc} mean-field 
expression to fit our simulation data to provide some insight into the role of attractive 
crowder-protein interactions in destabilizing protein association. 
However, there is a need for comprehensive quantitative theory, that can 
describe the effects of repulsive as well as attractive crowder-protein 
interactions on the protein-association equilibria.

In this paper, we present a theory that can quantitatively predict the effects of
macromolecular crowding on the
protein association equilibria accounting for both repulsive and attractive
crowder-protein interactions. The statistical mechanics and thermodynamics
of a hard-sphere fluid are adapted to yield an approximate analytical
expression for the protein-binding free energy in the presence of spherical
crowders. Extensive replica exchange Monte Carlo (REMC)
simulations have been performed on two distinct protein complexes to
test this theory. We find that the theory is in excellent agreement
with simulations over a wide range of crowder packing fractions
and crowder-protein interactions. The theory identifies the region in 
parameter space (entropy-enthalpy compensation line in a two parameter plane) separating 
entropically stabilized area versus enthalpically destabilized one.

\begin{figure}[t]
\includegraphics*[width=0.9 \columnwidth]{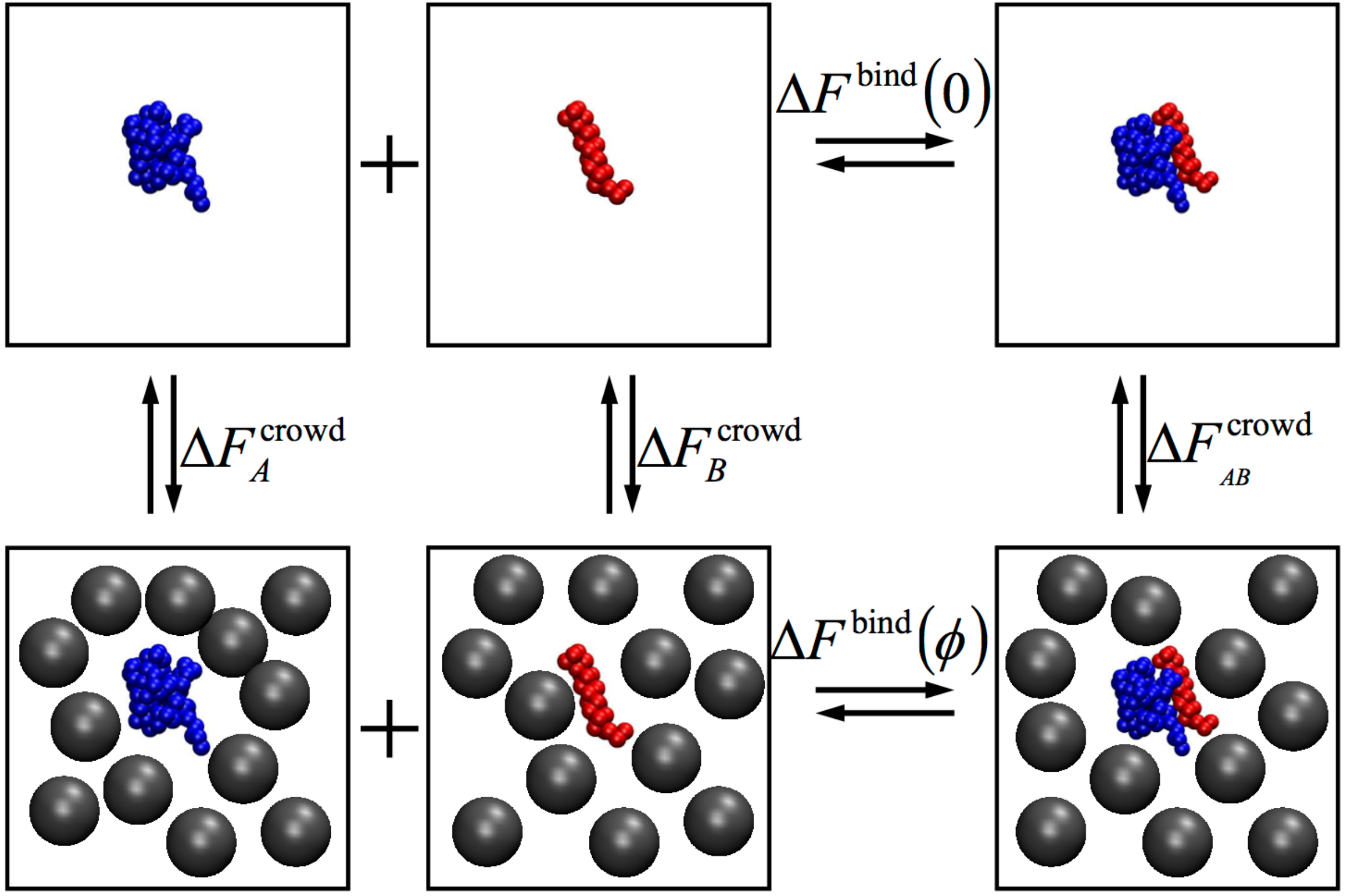}
\caption{\label{fig1} Schematic diagram of the thermodynamic cycle for
  the formation of the Ubq/UIM1 complex. The ubiquitin is shown in blue
  while UIM1 is shown in red.}
\end{figure}

{\em Theoretical development.}
Figure~\ref{fig1} illustrates a thermodynamic cycle that describes a
change in the binding free energy $\Delta F^{\mathrm {bind}}$ of two proteins due to the presence of
crowding molecules. This change, $\Delta\Delta F^{\mathrm {bind}}$, can be expressed
as the difference in the binding free energy 
in the absence and presence of crowders and is given by
\begin{eqnarray}
\label{eq:ddF_b}
\Delta\Delta F^{\mathrm {bind}}(\phi) &=& \Delta F^{\mathrm {bind}}(\phi) - \Delta F^{\mathrm {bind}}(\phi=0) \nonumber \\
                 &=& \Delta F_\text{AB}^\text{crowd} -
\Delta F_\text{A}^\text{crowd} - \Delta
F_\text{B}^\text{crowd},
\end{eqnarray}
where $\Delta F_\alpha^\text{crowd}(\phi), (\alpha \in$ [A,B,AB]) is the solvation
free energy of a protein (or complex) $\alpha$ in a crowded solution
with crowding packing fraction $\phi$. [For brevity, we will omit the superscript
  ``crowd'' below.]

To obtain an expression for $\Delta F_\alpha (\phi)$ in
Eq.~(\ref{eq:ddF_b}) for a protein or complex $\alpha$, let
$U_\alpha(r,\Omega) = \sum_{i\in\alpha} u_i(r_i)$ be the overall
interaction between a protein $\alpha$ and a crowder, where $r$ is
the distance between the center of mass of the protein and the crowder
and $\Omega$ the orientational degree of freedom, while $u_i$ is
the interaction between an atom (or residue) $i$ of the protein
$\alpha$ and the crowder. 
For a general Lennard-Jones(LJ)-type
potential for $u_i$, it is reasonable to assume that for given $\Omega$, $U_\alpha(r,\Omega)$
exhibits a minimum, $-\epsilon_\alpha^m(\Omega)$, at $r=r_\alpha^m(\Omega)$.
Following the Weeks-Chandler-Andersen (WCA) theory, we then decompose
$U_\alpha$ into the repulsive and attractive parts as
\begin{eqnarray}
\label{eq:U-WCA}
U_{\alpha,\text{rep}}(r,\Omega) & = & \left\{ \begin{array}{l l}
   U_\alpha(r,\Omega) + \epsilon_{\alpha}^ m(\Omega) & \quad \text{$r <
     r_{\alpha}^m(\Omega)$}, \\
   0  &  \quad  \text{otherwise},
  \end{array} \right. \nonumber \\
U_{\alpha,\text{att}}(r,\Omega) & = & \left\{ \begin{array}{l l}
   -\epsilon_{\alpha}^m(\Omega)  &  \quad  \text{$r < r_{\alpha}^m(\Omega)$}, \\
   U_\alpha(r,\Omega)  &  \quad  \text{otherwise}.
  \end{array} \right.
\end{eqnarray}

The solvation free energy, $\Delta F_\alpha(\phi)$, of the protein in
a crowded solution
can then be divided into two parts as, 
\begin{equation}
\label{eq:dF}
\Delta F_\alpha(\phi) = \Delta F_\text{$\alpha$,rep}(\phi) + \Delta
F_\text{$\alpha$,att}(\phi),
\end{equation}
where $\Delta F_\text{$\alpha$,rep(att)}$ is the contribution from the
repulsive (attractive) interaction, respectively.

The repulsive contribution, $\Delta F_\text{$\alpha$,rep}$, is
obtained by adopting the SPT. The SPT 
theory provides the
free energy for solvating a hard-sphere of radius
$R_\alpha$ in a bath of hard-sphere particles of radius
$R_c$ as, 
\begin{eqnarray}
\label{eq:dF_rep}
\Delta F_\text{$\alpha$,rep} & = & (3y + 3y^2 + y^3)\tilde{\phi} +
(4.5y^2 + 3y^3)\tilde{\phi}^2 \nonumber \\ 
&+& 3y^3\tilde{\phi}^3 - \ln (1-\phi),
\end{eqnarray}
where $\tilde{\phi} = \phi/(1-\phi)$ and $y=R_\alpha/R_c$. But can
we represent an anisometric protein with soft-core protein-crowder interactions as a 
hard sphere with an appropriate radius 
$R_{\alpha}$ to capture protein's solvation behavior accurately? 
Here we use the Boltzmann criteria to
define $R_\alpha$ as,
\begin{equation}
\label{eq:R_alpha}
\frac{4\pi}{3} (R_\alpha + R_c)^3 = \int_{U_{\alpha,\text{rep}} = fk_\text{B}T}
r^2drd\Omega,
\end{equation}
where the right-hand side
represents the volume encompassed by the condition $U_{\alpha,\text{rep}}(r,\Omega)
\geq fk_\text{B}T$. Here, we use $f=2$ that has been used successfully
in previous studies~\cite{Mittal:2007p530}.

Using thermodynamic perturbation theory approach, the attractive
contribution, $\Delta F_\text{$\alpha$,att}$, can be expressed as (up to
the first order), 
\begin{equation}
\label{eq:dF_att}
\Delta F_\text{$\alpha$,att} \approx \langle U_{\alpha,\text{att}}\rangle_\text{rep} = \int\rho
U_\text{$\alpha$,att}(r,\Omega)g_0(r) r^2 drd\Omega,
\end{equation}
where $\rho$ is the
crowder number density related to $\phi$ via $\rho
= \phi/(4\pi R_c^3/3)$, and
$g_0(r)$ is the radial distribution function of the hard-sphere
crowders between a protein and a crowder. Realizing that $g_0(r)$ has
a maximum $g_{0}^{\mathrm {max}}$ at contact and then
decays rapidly to unity, we assume $g_0(r)=g_{0}^{\mathrm {max}}$ for $r\in
[r_\alpha^m,r_\alpha^m+\lambda)$ and $1$ for $r\in
  [r_\alpha^m+\lambda,\infty)$ with $\lambda = (2^{1/6}-1)R_c \simeq
    0.12R_c$~\cite{Garde-BC-1999}. We then approximate Eq.(\ref{eq:dF_att}) as, 
\begin{equation}
\label{eq:dF_att_approx}
\Delta F_{\alpha,\text{att}} \approx -\rho \bar{\epsilon}_\alpha S_\alpha
\{\delta r + (g_{0}^{\mathrm {max}}-1)\lambda\},
\end{equation}
where $\bar{\epsilon}_\alpha = \langle \epsilon_\alpha^m \rangle_\Omega$ is
the orientational average of $\epsilon_\alpha^m$,
$S_\alpha=\int [r_\alpha^m(\Omega)]^2d\Omega$ the surface area around the protein, and
$\delta r$ the attraction range. Note that here we assume $\delta r \geq \lambda$.

To enhance the simplicity and practical value of our theory, we use
the Carnahan-Starling (CS) equation of state for a hard sphere
fluid to calculate $g_{0}^{\mathrm {max}}$.  The CS equation of state is known to reproduce 
the thermodynamic behavior of hard-sphere fluids from dilute gas to  
near the freezing 
transition. The CS expression for $g_{0}^{\mathrm {max}}$ is given by 
\begin{equation}
\label{eq:gm-CS}
g_{0}^{\mathrm {max}} = g_{\mathrm {CS}}^\text{max}(\phi) = (1-\phi/2)/(1-\phi)^3,
\end{equation}
 and only depends on $\phi$. Note that the first term in
 Eq.~(\ref{eq:dF_att_approx}) gives a linear order in $\phi$ while the
 term containing $g_{0}^{\mathrm {max}}$ yields higher order terms. Combining together 
Eqs.~(1), (3), (4) ,(7) and (8), one can easily obtain an estimate of crowding induced change in 
the binding free energy.
Next, we test this theory against REMC simulations of two protein
complexes in a wide range of crowder sizes, packing fractions and
interaction strengths.

{\em Model and simulation details.}
A residue-based coarse-grained model is used to simulate
protein-protein interactions~\cite{Kim-JMB-2008}. This transferable protein-protein 
interaction model was shown to yield binding
affinities and structures for moderate-to-weakly interacting protein
complexes in accord with experiments~\cite{Kim-JMB-2008,Kim-PNAS-2008}. 
Crowding agents are represented by
spheres interacting via a repulsive potential, $u_\text{rep}(r) =
\epsilon_r\left(\frac{\sigma_r}{r - 2r_c +
  \sigma_r}\right)^{12}$, where $\sigma_r$ is the interaction range set
equal to 6\AA. As our protein-protein interaction model only includes solvent (water)  
effects indirectly by accounting for it in the amino acid pair contact potentials, repulsive 
crowder-crowder interactions essentially mean that crowder-solvent interactions are assumed to 
be much stronger (to keep crowders dispersed in solution). 
See Supplemental 
Material (SM) at [URL will be inserted by publisher] for more details on models 
and simulation.


\begin{figure}[t]
\includegraphics*[width=0.8\columnwidth]{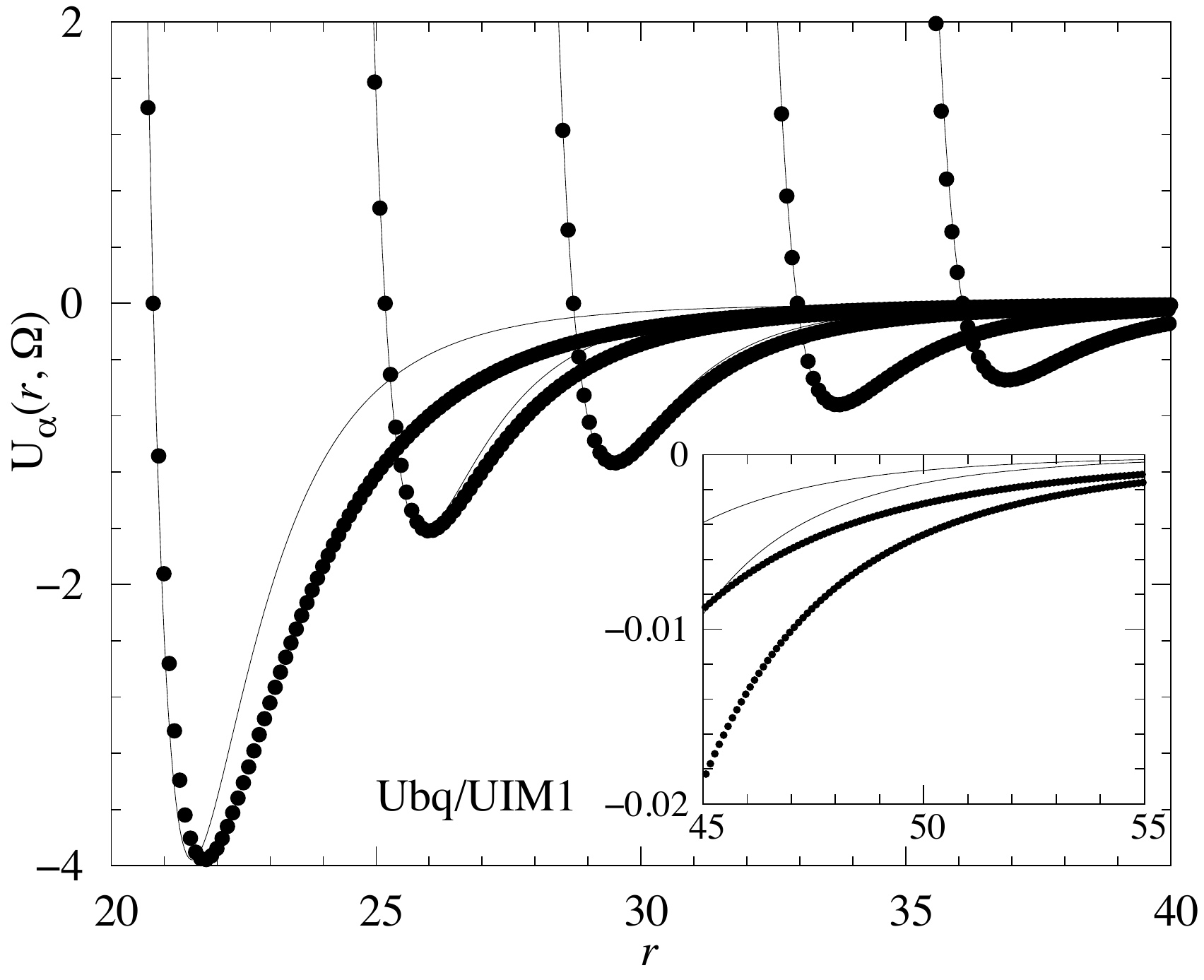}
\caption{\label{fig2} Plot of the overall interaction between the
  Ubq/UIM1 complex and a crowder, $U_\text{Ubq/UIM1}$, as a function of $r$ for
  different orientations $\Omega$. The solid curves are obtained from
  SM Eq. (1) with $\epsilon_c=\epsilon_\text{min}(\Omega)$
and $\sigma_i=r_0(\Omega)$.}
\end{figure}

{\em Results and Discussion.}
The spherical crowders interact with each other via the distance-dependent soft 
repulsive potential given by $u_\text{rep}(r)$ with a characteristic size $r_c$
and $\epsilon_r = 1.69 k_\text{B}T$. 
To apply the SPT theory (4) for
calculating the repulsive contribution of the binding free energy, it
is necessary to obtain an effective hard-sphere radius for such
crowders. We define the effective hard-sphere radius,
$R_c$, of crowders by
the condition, $u_\text{rep}(2R_c) = fk_\text{B}T$ with the same $f$
as in Eq.~(\ref{eq:R_alpha}). This yields $R_c = r_c + \gamma \sigma_r$ where $\gamma =
\frac{1}{2}[(\frac{1.69}{2.0})^{1/12}-1]$. Note that although for $\epsilon_r =
1.69 k_\text{B}T$ one has $R_c \simeq r_c$, in general, $R_c$ can be different from $r_c$. 
The effective packing fraction $\phi$ is then given by $\phi = \phi_0 (R_c/r_c)^{3}$. 

Figure~\ref{fig2} presents the overall interaction between
the complex Ubq/UIM1 and a crowder at
five different orientations, illustrating a highly anisotropic and
asymmetric nature of the interaction. It shows that the overall protein-crowder
interaction follows the LJ shape of the residue-crowder interaction,
SM Eq. 1, (see the solid curves), with a minimum
$-\epsilon_m(\Omega)$ at $r = r_m(\Omega)$ for a given $\Omega$. However, the longer-distance tails are
underestimated by the same formula as evident in the inset.

\begin{figure}
\includegraphics*[width=0.9\columnwidth]{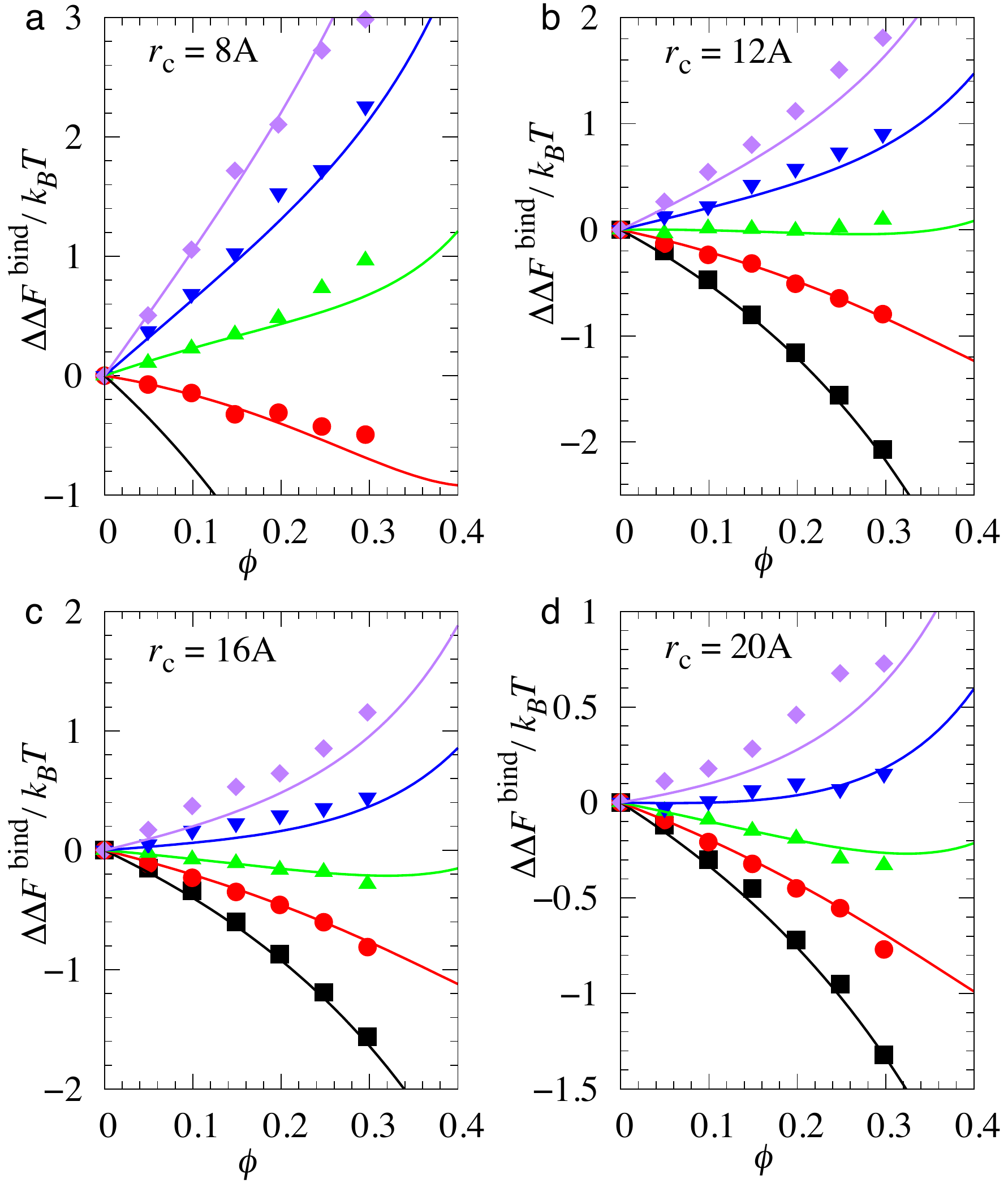}
\caption{\label{fig3} Binding free energy, $\Delta\Delta F_b(\phi)$,
  for the Ubq/UIM1 complex as a function of the crowder
  packing fraction $\phi$. The symbols and solid curves (black: $\epsilon_r$ = 1.69 $k_{\mathrm B}T$ for 
repulsive interactions, red: $\epsilon_c$ = 0.15 $k_{\mathrm B}T$, green: 0.3 $k_{\mathrm B}T$, blue: 0.45 $k_{\mathrm B}T$, purple: 0.6 $k_{\mathrm B}T$ for attractive interactions) are simulation data and predictions 
from the theory, respectively (see the text).}
\end{figure}

The effective radius, $R_\alpha$, for a
protein $\alpha$, determined by Eq.~(\ref{eq:R_alpha}) depends weakly on
$r_c$ and $\epsilon_c$ (see SM) as shown in Table~\ref{tab1}. For the repulsive
protein-crowder interactions, such effective
radii for proteins and complexes are sufficient enough to
calculate the change in the binding
free energy, $\Delta\Delta F^{\mathrm {bind}}$, via
Eq.~(\ref{eq:dF_rep}). Figure~\ref{fig3} shows an excellent agreement
between simulation results (black squares) and the theory (black solid
curves) for the Ubq/UIM1 complex for different crowder sizes. As
previously reported by us and others, the binding free energy decreases with increasing packing fraction
$\phi$ and decreasing crowder size due to the excluded-volume effect.

\begingroup
\squeezetable
\begin{table}
\caption{\label{tab1} Effective radius, $R_\alpha$, (in \AA), for the ubiquitin (Ubq),
  UIM1 and the Ubq/UIM1 complex for $r_c = 12, 16, 20$\AA for
  attractive ($\epsilon_c = 0.15, 0.3, 0.45, 0.6 k_\text{B}T$) and repulsive (rep;
$\epsilon_r = 1.69 k_\text{B}T$) interactions}
\begin{ruledtabular}
\begin{tabular}{c|ccc|ccc|ccc}
  & \multicolumn{3}{c|}{Ubq}  &
 \multicolumn{3}{c|}{UIM1} & \multicolumn{3}{c}{Ubq/UIM1}
 \\
 $\epsilon_c$ & 12 & 16 & 20 & 12 & 16 & 20 & 12 & 16 & 20 \\ \hline
 0.15 & 14.13 & 14.39 & 14.57 & 9.82 & 10.14 & 10.37 & 15.88 & 16.17 & 16.38  \\ \hline
 0.30 & 14.29 & 14.54 & 14.72 & 9.99 & 10.31 & 10.54 & 16.03 & 16.32 & 16.53  \\ \hline
 0.45 & 14.36 & 14.61 & 14.79 & 10.07 & 10.38 & 10.62 & 16.11 & 16.39 & 16.60  \\ \hline
 0.60 & 14.41 & 14.65 & 14.83 & 10.12 & 10.43 & 10.66 & 16.15 & 16.44
 & 16.64  \\ \hline
 rep  & 15.18 & 15.42 & 15.59 & 10.83 & 11.13 & 11.35 & 16.92 & 17.20 & 17.40 \\

\end{tabular}
\end{ruledtabular}
\end{table}
\endgroup

Recent studies~\cite{Douglas-PRL-2009,Jiao-BPJ-2010,Rosen-JPC-2011} have shown that
attractive protein-crowder interactions can destabilize 
protein association. Figure~\ref{fig3} shows that \textit{indeed} as
the attraction strength, $\epsilon_c$, between a residue and a crowder
increases the binding free energy also increases with the packing
fraction $\phi$. For example, for a moderate strength $\epsilon_c =
0.6k_\text{B}T$ the change in the binding free energy at $\phi = 0.3$
(close to the physiological condition) is up to about 4 $k_\text{B}T$ when
the protein-crowder interaction switches from repulsive (black) to
attractive (purple). For reference, hard sphere fluids undergo freezing 
transition at $\phi = 0.49$ and the random close packing is 
$\phi = 0.64$~\cite{torq}. 
After including the volume occupied by the proteins, it is clear that we are 
not simulating low crowder packing fractions for which linear expansion in 
$\phi$ can explain the observed trends.   
In order to apply our theory,
Eqs.~(\ref{eq:ddF_b})-(\ref{eq:gm-CS}), to describe the simulation
data for various $\epsilon_c$ and $r_c$, we calculate the average
attraction strength, $\bar{\epsilon}_\alpha$, and the surface area,
$S_\alpha$, for the individual proteins and the
complex. Note that $\bar{\epsilon}_\alpha$ is proportional to
$\epsilon_c$ while $S_\alpha$ is independent of
$\epsilon_c$. Table~\ref{tab2} shows these values for different $r_c$. The theory 
predictions are in excellent agreement with
the simulation data in which the attraction range $\delta r = 5$ \AA
(close to $\sigma_r$) is used for all the crowder sizes and
attraction strengths.

\begin{table}[b]
\caption{\label{tab2} Normalized average attraction strength,
  $\bar{\epsilon}_\alpha/\epsilon_c$, and the surface area,
  $S_\alpha$, (in \AA$^3$) for Ubq,
  UIM1 and the Ubq/UIM1 complex}
\begin{ruledtabular}
\begin{tabular}{c|cc|cc|cc}
   & \multicolumn{2}{c|}{Ubq}  &
 \multicolumn{2}{c|}{UIM1} & \multicolumn{2}{c}{Ubq/UIM1}
 \\
 $r_c$ &  $\bar{\epsilon}_\alpha/\epsilon_c$ & $S_\alpha$ & $\bar{\epsilon}_\alpha/\epsilon_c$ & $S_\alpha$ & $\bar{\epsilon}_\alpha/\epsilon_c$ & $S_\alpha$ \\ \hline
 8 & 4.56 & 6543 &  4.01 & 4100 & 4.61 & 7522  \\ \hline
 12 & 4.71 & 9278 &  4.07 & 6418 & 4.75 & 10487  \\ \hline
 16 & 4.79 & 12402 & 4.07 & 9143 & 4.85 & 13830  \\ \hline
 20 & 4.85 & 15921 & 4.04 & 12273 & 4.91 & 17562  \\

\end{tabular}
\end{ruledtabular}
\end{table}

\begin{figure}
\includegraphics*[width=0.9\columnwidth]{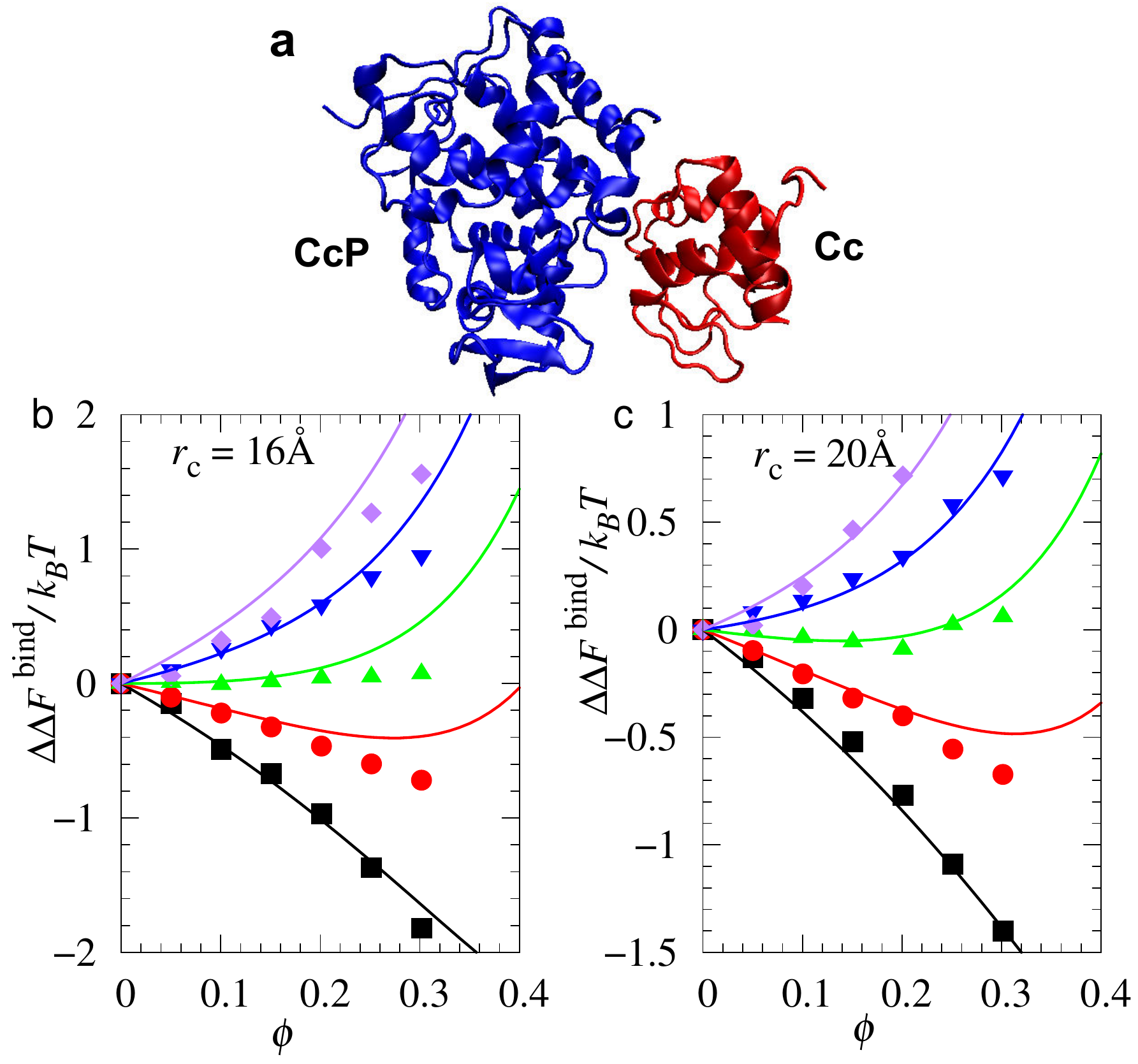}
\caption{\label{fig4} Binding free energy, $\Delta\Delta F_b$, for the
  Cc/CcP complex as a function of $\phi$. Symbols and curves are same
  as in Fig.~3.}
\end{figure}

To check whether the theory can be transferable to other
protein complexes, we calculate the binding free energies
for the Cc/CcP complex (total 402 residues compared to 100 residues
for the Ubq/UIM1 complex) as shown in Fig.~\ref{fig4}. With the same
$\delta r$, the theoretical predictions agree remarkably well with the
simulation data. 

The data in Figures 3 and 4 show the competition between
entropic effects of the excluded volume and enthalpic effects by
attractive crowder-protein interactions. As previously
suggested~\cite{Jiao-BPJ-2010,Rosen-JPC-2011}, the enthalpic effects
can be approximated to be proportional to protein's surface areas and our theory
here provides its concrete foundation from microscopic nature of the
protein-crowder interactions. At high attraction strengths, 
the enthalpic penalty for breaking the crowder-protein interactions (at the 
expense of protein-protein interactions)
dominates, thus increasing the binding free energy. At some critical
attraction $\epsilon_c^\text{crit}$, the two contributions are canceled out, and
the binding energy in a crowded solution becomes equal to that in the
absence of crowders (see green triangles and curve in Fig.~3b).

\begin{figure}
\includegraphics*[width=0.8\columnwidth]{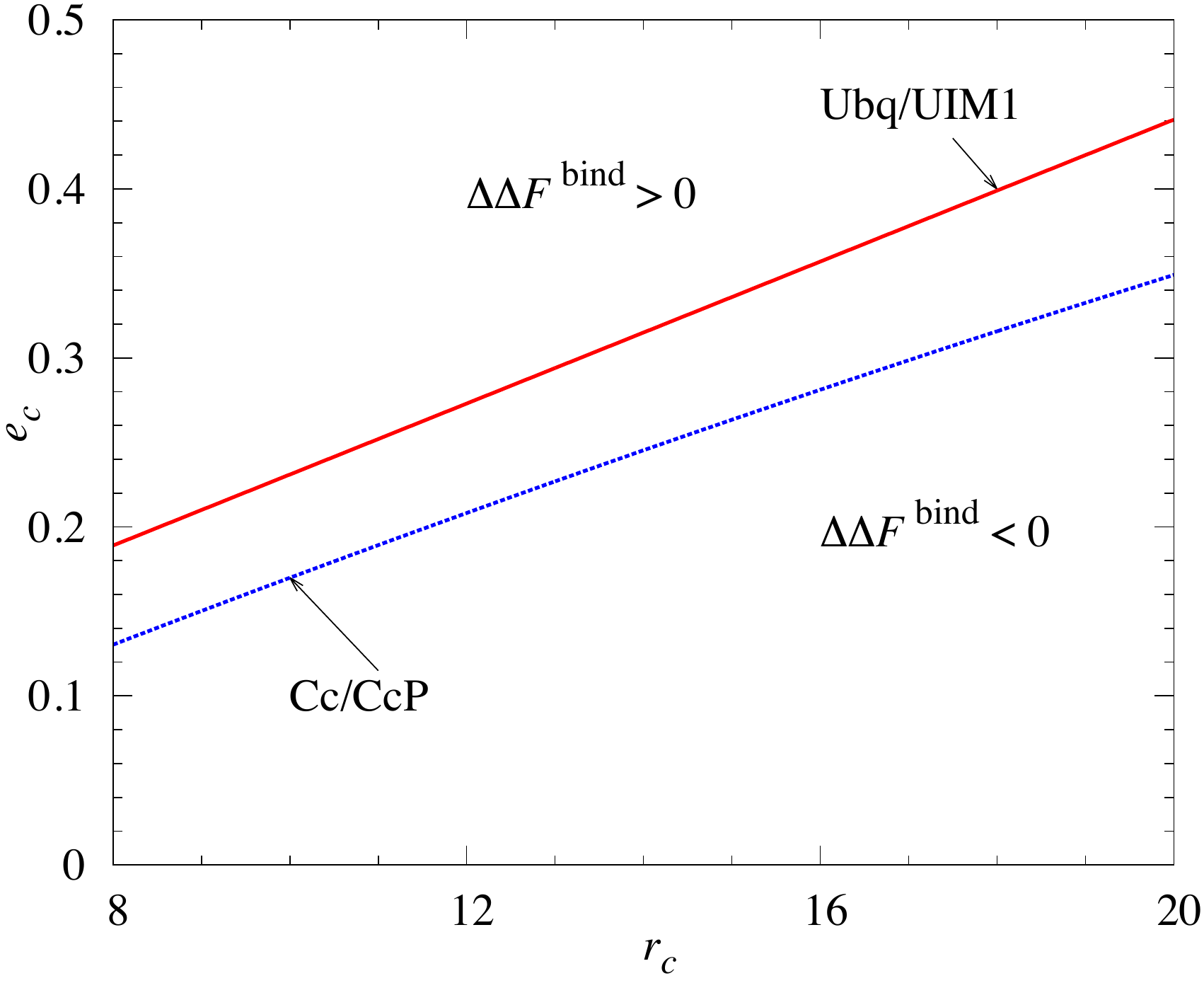}
\caption{\label{fig5} Enthalpy-entropy compensation lines (i.e.,
  $\Delta\Delta F^\text{bind} = 0$) in the
  parameter space $(\epsilon_c,r_c)$ for Ubq/UIM1 and Cc/CcP complexes.}
\end{figure}

It was observed \cite{Rosen-JPC-2011} that the critical attraction,
$\epsilon_c^\text{crit}$, for which the effect of the excluded volume is
canceled out exactly by that of the attractive contribution, (i.e.,
$\Delta\Delta F^{\mathrm {bind}} = 0$), is approximately independent of the crowder 
packing fraction, $\phi$. This is owing to the fact that $\Delta\Delta F^{\mathrm {bind}}$ is
almost linear in $\phi$ for $\epsilon_c$ considered. To obtain
$\epsilon_c^\text{crit}$ estimate, we combine Eqs.~(\ref{eq:dF_rep}) and
(\ref{eq:dF_att_approx}) and solve for $\epsilon_c$ that satisfies $\Delta\Delta F^{\mathrm {bind}} =
0$ up to the linear order in $\phi$. One then obtains, 
\begin{equation}
\label{eq:e_crit}
\epsilon_c^\text{crit} = \Delta Y/ \Delta W + O(\phi),
\end{equation}
where
\begin{eqnarray}
\label{eq:DY}
\Delta Y & = & 3(y_A+y_B-y_{AB}) + 3(y_A^2 + y_B^2 - y_{AB}^2)
\nonumber \\
  &  &+ (y_A^3 + y_B^3 - y_{AB}^3) + 1, \\
\Delta W & = & 3(\bar{\epsilon}_AS_A + \bar{\epsilon}_BS_B -
\bar{\epsilon}_{AB}S_{AB})\delta r/(4\pi R_c^3 e_c).
\end{eqnarray}
This yields $\epsilon_c^\text{crit}/k_\text{B}T \simeq 0.19$, 0.27, 0.36 and 0.44 for $r_c = 8,
12, 16, 20$\AA, for the Ubq/UIM1, and 0.28 and 0.35 for $r_c=16$ and 20\AA for
the Cc/CcP, respectively, consistent with simulation data in Figs.~\ref{fig3} and \ref{fig4}.
We can also plot $\epsilon_c^\text{crit}$ as it changes with crowder size $r_c$ as shown in 
Figure~\ref{fig5}. For crowder-protein attraction values above this line, one will 
observe destabilization of protein association and stabilization below this line.

In summary, we have presented a quantitative theory for 
protein association equilibria in a crowded solution for both repulsive
and attractive crowder-protein interactions. 
This work is important for providing a theoretical foundation of understanding the
protein-protein interactions in a cellular environment in which
proteins and crowding macromolecules exhibit non-specific interactions in
addition to the excluded volume effects.
The theory is based on the
statistical mechanics and thermodynamics of a hard-sphere fluid. Even
though proteins are highly anisometric, the repulsive contribution to
the binding free energy is described well by the scaled particle
theory of hard spheres. The expression for the attractive contribution
is obtained by using thermodynamic perturbation theory and the radial
distribution function of hard-sphere fluids. The theory is in excellent agreement with
simulation results for the Ubq/UIM1 and Cc/CcP complexes over a wide
range of the crowder sizes, packing fractions and attraction
strengths. 

We also observe crowding induced compensation for a critical protein-crowder interaction 
strength (indpendent of crowder packing fraction) leading to no change in the  
binding free energy with respect to bulk. Earlier Trout and co-workers had 
proposed a neutral-crowder hypothesis to explain the kinetic effect of small 
solution additives (crowders) that slow down the rate of protein association and 
dissociation without perturbing the equilibrium~\cite{baynes2004rational}. 
It will be interesting, in future, 
to explore the protein kinetics near this critial protein-crowder interaction strength 
to test if the neutral-crowder hypothesis is applicable in general. In future, we also 
plan to include attractions in the crowder-crowder interaction potential to study their 
interplay with protein-crowder and protein-protein interactions~\cite{kim2010crowding}. 


\begin{thebibliography}{38}
\expandafter\ifx\csname natexlab\endcsname\relax\def\natexlab#1{#1}\fi
\expandafter\ifx\csname bibnamefont\endcsname\relax
  \def\bibnamefont#1{#1}\fi
\expandafter\ifx\csname bibfnamefont\endcsname\relax
  \def\bibfnamefont#1{#1}\fi
\expandafter\ifx\csname citenamefont\endcsname\relax
  \def\citenamefont#1{#1}\fi
\expandafter\ifx\csname url\endcsname\relax
  \def\url#1{\texttt{#1}}\fi
\expandafter\ifx\csname urlprefix\endcsname\relax\def\urlprefix{URL }\fi
\providecommand{\bibinfo}[2]{#2}
\providecommand{\eprint}[2][]{\url{#2}}

\bibitem[{\citenamefont{Kleanthous}(2000)}]{Kleanthous-2000}
\bibinfo{author}{\bibfnamefont{C.}~\bibnamefont{Kleanthous}},
  \emph{\bibinfo{title}{Protein-Protein Recognition}}
  (\bibinfo{publisher}{Oxford University Press}, \bibinfo{address}{Oxford, UK},
  \bibinfo{year}{2000}).

\bibitem[{\citenamefont{Tang et~al.}(2006)\citenamefont{Tang, Iwahara, and
  Clore}}]{Tang-N-2006}
\bibinfo{author}{\bibfnamefont{C.}~\bibnamefont{Tang}},
  \bibinfo{author}{\bibfnamefont{J.}~\bibnamefont{Iwahara}}, \bibnamefont{and}
  \bibinfo{author}{\bibfnamefont{G.~M.} \bibnamefont{Clore}},
  \bibinfo{journal}{Nature} \textbf{\bibinfo{volume}{444}},
  \bibinfo{pages}{383} (\bibinfo{year}{2006}).

\bibitem[{\citenamefont{Boehr et~al.}(2008)\citenamefont{Boehr, Wright
  et~al.}}]{boehr2008}
\bibinfo{author}{\bibfnamefont{D.}~\bibnamefont{Boehr}},
  \bibinfo{author}{\bibfnamefont{P.}~\bibnamefont{Wright}},
  \bibnamefont{et~al.}, \bibinfo{journal}{Science}
  \textbf{\bibinfo{volume}{320}}, \bibinfo{pages}{1429} (\bibinfo{year}{2008}).

\bibitem[{\citenamefont{Elcock et~al.}(1999)\citenamefont{Elcock, Gabdoulline,
  Wade, and McCammon}}]{elcock1999}
\bibinfo{author}{\bibfnamefont{A.}~\bibnamefont{Elcock}},
  \bibinfo{author}{\bibfnamefont{R.}~\bibnamefont{Gabdoulline}},
  \bibinfo{author}{\bibfnamefont{R.}~\bibnamefont{Wade}}, \bibnamefont{and}
  \bibinfo{author}{\bibfnamefont{J.}~\bibnamefont{McCammon}},
  \bibinfo{journal}{Journal of Molecular Biology}
  \textbf{\bibinfo{volume}{291}}, \bibinfo{pages}{149} (\bibinfo{year}{1999}).

\bibitem[{\citenamefont{Gilson and Zhou}(2007)}]{gilson2007}
\bibinfo{author}{\bibfnamefont{M.}~\bibnamefont{Gilson}} \bibnamefont{and}
  \bibinfo{author}{\bibfnamefont{H.-X.} \bibnamefont{Zhou}},
  \bibinfo{journal}{Annu. Rev. Biophys. Biomol. Struct.}
  \textbf{\bibinfo{volume}{36}}, \bibinfo{pages}{21} (\bibinfo{year}{2007}).

\bibitem[{\citenamefont{Fulton}(1982)}]{Fulton:1982p8511}
\bibinfo{author}{\bibfnamefont{A.}~\bibnamefont{Fulton}},
  \bibinfo{journal}{Cell} \textbf{\bibinfo{volume}{30}}, \bibinfo{pages}{345}
  (\bibinfo{year}{1982}).

\bibitem[{\citenamefont{Zimmerman and Minton}(1993)}]{Zimmerman:1993p2187}
\bibinfo{author}{\bibfnamefont{S.}~\bibnamefont{Zimmerman}} \bibnamefont{and}
  \bibinfo{author}{\bibfnamefont{A.}~\bibnamefont{Minton}},
  \bibinfo{journal}{Annual Review of Biophysics and Biomolecular Structure}
  \textbf{\bibinfo{volume}{22}}, \bibinfo{pages}{27} (\bibinfo{year}{1993}).

\bibitem[{\citenamefont{Cheung et~al.}(2005)\citenamefont{Cheung, Klimov, and
  Thirumalai}}]{Cheung-2005}
\bibinfo{author}{\bibfnamefont{M.~S.} \bibnamefont{Cheung}},
  \bibinfo{author}{\bibfnamefont{D.}~\bibnamefont{Klimov}}, \bibnamefont{and}
  \bibinfo{author}{\bibfnamefont{D.}~\bibnamefont{Thirumalai}},
  \bibinfo{journal}{Proc. Natl. Acad. Sci. USA} \textbf{\bibinfo{volume}{102}},
  \bibinfo{pages}{4753} (\bibinfo{year}{2005}).

\bibitem[{\citenamefont{Zhou et~al.}(2008)\citenamefont{Zhou, Rivas, and
  Minton}}]{Zhou-ARB-2008}
\bibinfo{author}{\bibfnamefont{H.-X.} \bibnamefont{Zhou}},
  \bibinfo{author}{\bibfnamefont{G.}~\bibnamefont{Rivas}}, \bibnamefont{and}
  \bibinfo{author}{\bibfnamefont{A.~P.} \bibnamefont{Minton}},
  \bibinfo{journal}{Annu. Rev. Biophys.} \textbf{\bibinfo{volume}{37}},
  \bibinfo{pages}{375} (\bibinfo{year}{2008}).

\bibitem[{\citenamefont{Shen et~al.}(2009)\citenamefont{Shen, Cheung,
  Errington, and Truskett}}]{Shen-2009}
\bibinfo{author}{\bibfnamefont{V.~K.} \bibnamefont{Shen}},
  \bibinfo{author}{\bibfnamefont{J.~K.} \bibnamefont{Cheung}},
  \bibinfo{author}{\bibfnamefont{J.~R.} \bibnamefont{Errington}},
  \bibnamefont{and} \bibinfo{author}{\bibfnamefont{T.~M.}
  \bibnamefont{Truskett}}, \bibinfo{journal}{J. Biomech. Eng.}
  \textbf{\bibinfo{volume}{131}}, \bibinfo{pages}{071002}
  (\bibinfo{year}{2009}).

\bibitem[{\citenamefont{Kim and Yethiraj}(2009)}]{Kim-BPJ-2009}
\bibinfo{author}{\bibfnamefont{J.~S.} \bibnamefont{Kim}} \bibnamefont{and}
  \bibinfo{author}{\bibfnamefont{A.}~\bibnamefont{Yethiraj}},
  \bibinfo{journal}{Biophys. J.} \textbf{\bibinfo{volume}{96}},
  \bibinfo{pages}{1333} (\bibinfo{year}{2009}).

\bibitem[{\citenamefont{Elcock}(2010)}]{Elcock-COSB-2010}
\bibinfo{author}{\bibfnamefont{A.~H.} \bibnamefont{Elcock}},
  \bibinfo{journal}{Curr. Opin. Struct. Biol.} \textbf{\bibinfo{volume}{20}},
  \bibinfo{pages}{196} (\bibinfo{year}{2010}).

\bibitem[{\citenamefont{Mittal and Best}(2010)}]{mittal-2010}
\bibinfo{author}{\bibfnamefont{J.}~\bibnamefont{Mittal}} \bibnamefont{and}
  \bibinfo{author}{\bibfnamefont{R.~B.} \bibnamefont{Best}},
  \bibinfo{journal}{Biophys. J.} \textbf{\bibinfo{volume}{98}},
  \bibinfo{pages}{315} (\bibinfo{year}{2010}).

\bibitem[{\citenamefont{Minton and Wilf}(1981)}]{minton-1981a}
\bibinfo{author}{\bibfnamefont{A.~P.} \bibnamefont{Minton}} \bibnamefont{and}
  \bibinfo{author}{\bibfnamefont{J.}~\bibnamefont{Wilf}},
  \bibinfo{journal}{Biochemistry} \textbf{\bibinfo{volume}{20}},
  \bibinfo{pages}{4821} (\bibinfo{year}{1981}).

\bibitem[{\citenamefont{Jarvis et~al.}(1990)\citenamefont{Jarvis, Ring, Daube,
  and von Hippel}}]{Jarvis-1990}
\bibinfo{author}{\bibfnamefont{T.~C.} \bibnamefont{Jarvis}},
  \bibinfo{author}{\bibfnamefont{D.~M.} \bibnamefont{Ring}},
  \bibinfo{author}{\bibfnamefont{S.~S.} \bibnamefont{Daube}}, \bibnamefont{and}
  \bibinfo{author}{\bibfnamefont{P.~H.} \bibnamefont{von Hippel}},
  \bibinfo{journal}{J. Biol. Chem.} \textbf{\bibinfo{volume}{265}},
  \bibinfo{pages}{15160} (\bibinfo{year}{1990}).

\bibitem[{\citenamefont{van~den Berg et~al.}(1999)\citenamefont{van~den Berg,
  Ellis, and Dobson}}]{berg-1999}
\bibinfo{author}{\bibfnamefont{B.}~\bibnamefont{van~den Berg}},
  \bibinfo{author}{\bibfnamefont{R.~J.} \bibnamefont{Ellis}}, \bibnamefont{and}
  \bibinfo{author}{\bibfnamefont{C.~M.} \bibnamefont{Dobson}},
  \bibinfo{journal}{EMBO J.} \textbf{\bibinfo{volume}{18}},
  \bibinfo{pages}{6927} (\bibinfo{year}{1999}).

\bibitem[{\citenamefont{Wenner and Bloomfield}(1999)}]{Wenner:2008p8635}
\bibinfo{author}{\bibfnamefont{J.~R.} \bibnamefont{Wenner}} \bibnamefont{and}
  \bibinfo{author}{\bibfnamefont{V.~A.} \bibnamefont{Bloomfield}},
  \bibinfo{journal}{Biophysical journal} \textbf{\bibinfo{volume}{77}},
  \bibinfo{pages}{3234} (\bibinfo{year}{1999}).

\bibitem[{\citenamefont{Morar et~al.}(2001)\citenamefont{Morar, Wang, and
  Pielak}}]{Morar:2001p8850}
\bibinfo{author}{\bibfnamefont{A.~S.} \bibnamefont{Morar}},
  \bibinfo{author}{\bibfnamefont{X.}~\bibnamefont{Wang}}, \bibnamefont{and}
  \bibinfo{author}{\bibfnamefont{G.~J.} \bibnamefont{Pielak}},
  \bibinfo{journal}{Biochemistry} \textbf{\bibinfo{volume}{40}},
  \bibinfo{pages}{281} (\bibinfo{year}{2001}).

\bibitem[{\citenamefont{Patel et~al.}(2002)\citenamefont{Patel, Noble,
  Weatherly, Tripathy, Winzor, and Pielak}}]{patel2002}
\bibinfo{author}{\bibfnamefont{C.}~\bibnamefont{Patel}},
  \bibinfo{author}{\bibfnamefont{S.}~\bibnamefont{Noble}},
  \bibinfo{author}{\bibfnamefont{G.}~\bibnamefont{Weatherly}},
  \bibinfo{author}{\bibfnamefont{A.}~\bibnamefont{Tripathy}},
  \bibinfo{author}{\bibfnamefont{D.}~\bibnamefont{Winzor}}, \bibnamefont{and}
  \bibinfo{author}{\bibfnamefont{G.}~\bibnamefont{Pielak}},
  \bibinfo{journal}{Protein science} \textbf{\bibinfo{volume}{11}},
  \bibinfo{pages}{997} (\bibinfo{year}{2002}).

\bibitem[{\citenamefont{Kozer and Schreiber}(2004)}]{kozer-2004}
\bibinfo{author}{\bibfnamefont{N.}~\bibnamefont{Kozer}} \bibnamefont{and}
  \bibinfo{author}{\bibfnamefont{G.}~\bibnamefont{Schreiber}},
  \bibinfo{journal}{J. Mol. Biol.} \textbf{\bibinfo{volume}{336}},
  \bibinfo{pages}{763} (\bibinfo{year}{2004}).

\bibitem[{\citenamefont{Zorrilla et~al.}(2004)\citenamefont{Zorrilla, Rivas,
  Acu{\~n}a, and Lillo}}]{Zorrilla:2004p8683}
\bibinfo{author}{\bibfnamefont{S.}~\bibnamefont{Zorrilla}},
  \bibinfo{author}{\bibfnamefont{G.}~\bibnamefont{Rivas}},
  \bibinfo{author}{\bibfnamefont{A.~U.} \bibnamefont{Acu{\~n}a}},
  \bibnamefont{and} \bibinfo{author}{\bibfnamefont{M.~P.} \bibnamefont{Lillo}},
  \bibinfo{journal}{Protein Sci.} \textbf{\bibinfo{volume}{13}},
  \bibinfo{pages}{2960} (\bibinfo{year}{2004}).

\bibitem[{\citenamefont{Phillip et~al.}(2009)\citenamefont{Phillip, Sherman,
  Haran, and Schreiber}}]{Phillip-BPJ-2009}
\bibinfo{author}{\bibfnamefont{Y.}~\bibnamefont{Phillip}},
  \bibinfo{author}{\bibfnamefont{E.}~\bibnamefont{Sherman}},
  \bibinfo{author}{\bibfnamefont{G.}~\bibnamefont{Haran}}, \bibnamefont{and}
  \bibinfo{author}{\bibfnamefont{G.}~\bibnamefont{Schreiber}},
  \bibinfo{journal}{Biophys. J.} \textbf{\bibinfo{volume}{97}},
  \bibinfo{pages}{875} (\bibinfo{year}{2009}).

\bibitem[{\citenamefont{Wang et~al.}(2011)\citenamefont{Wang, Zhuravleva, and
  Gierasch}}]{Wang-Biochemistry-2011}
\bibinfo{author}{\bibfnamefont{Q.}~\bibnamefont{Wang}},
  \bibinfo{author}{\bibfnamefont{A.}~\bibnamefont{Zhuravleva}},
  \bibnamefont{and} \bibinfo{author}{\bibfnamefont{L.~M.}
  \bibnamefont{Gierasch}}, \bibinfo{journal}{Biochemistry}
  \textbf{\bibinfo{volume}{50}}, \bibinfo{pages}{9225} (\bibinfo{year}{2011}).

\bibitem[{\citenamefont{Fodeke and Minton}(2011)}]{fodeke2011}
\bibinfo{author}{\bibfnamefont{A.}~\bibnamefont{Fodeke}} \bibnamefont{and}
  \bibinfo{author}{\bibfnamefont{A.}~\bibnamefont{Minton}},
  \bibinfo{journal}{J. Phys. Chem. B} \textbf{\bibinfo{volume}{115}},
  \bibinfo{pages}{11261} (\bibinfo{year}{2011}).

\bibitem[{\citenamefont{Minton}(1983)}]{Minton-1983}
\bibinfo{author}{\bibfnamefont{A.~P.} \bibnamefont{Minton}},
  \bibinfo{journal}{Mol Cell. Biochem.} \textbf{\bibinfo{volume}{55}},
  \bibinfo{pages}{119} (\bibinfo{year}{1983}).

\bibitem[{\citenamefont{Zhou}(2004)}]{Zhou:2004p4459}
\bibinfo{author}{\bibfnamefont{H.-X.} \bibnamefont{Zhou}}, \bibinfo{journal}{J.
  Mol. Recognit.} \textbf{\bibinfo{volume}{17}}, \bibinfo{pages}{368}
  (\bibinfo{year}{2004}).

\bibitem[{\citenamefont{Kim et~al.}(2010)\citenamefont{Kim, Best, and
  Mittal}}]{Kim-JCP-2010}
\bibinfo{author}{\bibfnamefont{Y.~C.} \bibnamefont{Kim}},
  \bibinfo{author}{\bibfnamefont{R.~B.} \bibnamefont{Best}}, \bibnamefont{and}
  \bibinfo{author}{\bibfnamefont{J.}~\bibnamefont{Mittal}},
  \bibinfo{journal}{J. Chem. Phys.} \textbf{\bibinfo{volume}{133}},
  \bibinfo{pages}{205101} (\bibinfo{year}{2010}).

\bibitem[{\citenamefont{Douglas et~al.}(2009)\citenamefont{Douglas, Dudowicz,
  and Freed}}]{Douglas-PRL-2009}
\bibinfo{author}{\bibfnamefont{J.~F.} \bibnamefont{Douglas}},
  \bibinfo{author}{\bibfnamefont{J.}~\bibnamefont{Dudowicz}}, \bibnamefont{and}
  \bibinfo{author}{\bibfnamefont{K.~F.} \bibnamefont{Freed}},
  \bibinfo{journal}{Phys. Rev. Lett.} \textbf{\bibinfo{volume}{97}},
  \bibinfo{pages}{875} (\bibinfo{year}{2009}).

\bibitem[{\citenamefont{Jiao et~al.}(2010)\citenamefont{Jiao, Li, Chen, Minton,
  and Liang}}]{Jiao-BPJ-2010}
\bibinfo{author}{\bibfnamefont{M.}~\bibnamefont{Jiao}},
  \bibinfo{author}{\bibfnamefont{H.-T.} \bibnamefont{Li}},
  \bibinfo{author}{\bibfnamefont{J.}~\bibnamefont{Chen}},
  \bibinfo{author}{\bibfnamefont{A.~P.} \bibnamefont{Minton}},
  \bibnamefont{and} \bibinfo{author}{\bibfnamefont{Y.}~\bibnamefont{Liang}},
  \bibinfo{journal}{Biophy. J.} \textbf{\bibinfo{volume}{99}},
  \bibinfo{pages}{914} (\bibinfo{year}{2010}).

\bibitem[{\citenamefont{Rosen et~al.}(2011)\citenamefont{Rosen, Kim, and
  Mittal}}]{Rosen-JPC-2011}
\bibinfo{author}{\bibfnamefont{J.}~\bibnamefont{Rosen}},
  \bibinfo{author}{\bibfnamefont{Y.~C.} \bibnamefont{Kim}}, \bibnamefont{and}
  \bibinfo{author}{\bibfnamefont{J.}~\bibnamefont{Mittal}},
  \bibinfo{journal}{J. Phys. Chem. B} \textbf{\bibinfo{volume}{115}},
  \bibinfo{pages}{2683} (\bibinfo{year}{2011}).

\bibitem[{\citenamefont{Torquato et~al.}(2000)\citenamefont{Torquato, Truskett, and
  Debenedetti}}]{torq}
\bibinfo{author}{\bibfnamefont{S.}~\bibnamefont{Torquato}},
  \bibinfo{author}{\bibfnamefont{T.~M.} \bibnamefont{Truskett}}, \bibnamefont{and}
  \bibinfo{author}{\bibfnamefont{P.~G.}~\bibnamefont{Debenedetti}},
  \bibinfo{journal}{Phys. Rev. Lett.} \textbf{\bibinfo{volume}{84}},
  \bibinfo{pages}{2064} (\bibinfo{year}{2000}).

\bibitem[{\citenamefont{Phillip et~al.}(2012)\citenamefont{Phillip, Kiss, and
  Schreiber}}]{Phillip-PNAS-2012}
\bibinfo{author}{\bibfnamefont{Y.}~\bibnamefont{Phillip}},
  \bibinfo{author}{\bibfnamefont{V.}~\bibnamefont{Kiss}}, \bibnamefont{and}
  \bibinfo{author}{\bibfnamefont{G.}~\bibnamefont{Schreiber}},
  \bibinfo{journal}{Proc. Natl. Acad. Sci. USA} \textbf{\bibinfo{volume}{109}},
  \bibinfo{pages}{1461} (\bibinfo{year}{2012}).

\bibitem[{\citenamefont{Lebowitz and Rowlinson}(1964)}]{Lebowitz-JCP-1964}
\bibinfo{author}{\bibfnamefont{J.~L.} \bibnamefont{Lebowitz}} \bibnamefont{and}
  \bibinfo{author}{\bibfnamefont{J.~S.} \bibnamefont{Rowlinson}},
  \bibinfo{journal}{J. Chem. Phys.} \textbf{\bibinfo{volume}{41}},
  \bibinfo{pages}{133} (\bibinfo{year}{1964}).

\bibitem[{\citenamefont{Mittal et~al.}(2007)\citenamefont{Mittal, Errington,
  and Truskett}}]{Mittal:2007p530}
\bibinfo{author}{\bibfnamefont{J.}~\bibnamefont{Mittal}},
  \bibinfo{author}{\bibfnamefont{J.}~\bibnamefont{Errington}},
  \bibnamefont{and} \bibinfo{author}{\bibfnamefont{T.}~\bibnamefont{Truskett}},
  \bibinfo{journal}{Journal of Physical Chemistry B}
  \textbf{\bibinfo{volume}{111}}, \bibinfo{pages}{10054}
  (\bibinfo{year}{2007}).

\bibitem[{\citenamefont{Garde et~al.}(1999)\citenamefont{Garde, Garc\'{i}a,
  Pratt, and Hummer}}]{Garde-BC-1999}
\bibinfo{author}{\bibfnamefont{S.}~\bibnamefont{Garde}},
  \bibinfo{author}{\bibfnamefont{A.~E.} \bibnamefont{Garc\'{i}a}},
  \bibinfo{author}{\bibfnamefont{L.~R.} \bibnamefont{Pratt}}, \bibnamefont{and}
  \bibinfo{author}{\bibfnamefont{G.}~\bibnamefont{Hummer}},
  \bibinfo{journal}{Biophys. Chem.} \textbf{\bibinfo{volume}{78}},
  \bibinfo{pages}{21} (\bibinfo{year}{1999}).

\bibitem[{\citenamefont{Kim and Hummer}(2008)}]{Kim-JMB-2008}
\bibinfo{author}{\bibfnamefont{Y.~C.} \bibnamefont{Kim}} \bibnamefont{and}
  \bibinfo{author}{\bibfnamefont{G.}~\bibnamefont{Hummer}},
  \bibinfo{journal}{J. Mol. Biol.} \textbf{\bibinfo{volume}{375}},
  \bibinfo{pages}{1416} (\bibinfo{year}{2008}).

\bibitem[{\citenamefont{Kim et~al.}(2008)\citenamefont{Kim, Tang, Clore, and
  Hummer}}]{Kim-PNAS-2008}
\bibinfo{author}{\bibfnamefont{Y.~C.} \bibnamefont{Kim}},
  \bibinfo{author}{\bibfnamefont{C.}~\bibnamefont{Tang}},
  \bibinfo{author}{\bibfnamefont{G.~M.} \bibnamefont{Clore}}, \bibnamefont{and}
  \bibinfo{author}{\bibfnamefont{G.}~\bibnamefont{Hummer}},
  \bibinfo{journal}{Proc. Natl. Acad. Sci. USA} \textbf{\bibinfo{volume}{105}},
  \bibinfo{pages}{12855} (\bibinfo{year}{2008}).

\bibitem[{\citenamefont{Baynes and Trout}(2004)}]{baynes2004rational}
\bibinfo{author}{\bibfnamefont{B.}~\bibnamefont{Baynes}} \bibnamefont{and}
  \bibinfo{author}{\bibfnamefont{B.}~\bibnamefont{Trout}},
  \bibinfo{journal}{Biophys. J.} \textbf{\bibinfo{volume}{87}},
  \bibinfo{pages}{1631} (\bibinfo{year}{2004}).

\bibitem[{\citenamefont{Kim and Yethiraj}(2010)}]{kim2010crowding}
\bibinfo{author}{\bibfnamefont{J.}~\bibnamefont{Kim}} \bibnamefont{and}
  \bibinfo{author}{\bibfnamefont{A.}~\bibnamefont{Yethiraj}},
  \bibinfo{journal}{The Journal of Physical Chemistry B}
  \textbf{\bibinfo{volume}{115}}, \bibinfo{pages}{347} (\bibinfo{year}{2010}).

\end{thebibliography}

\end{document}